\documentclass[acmsmall,screen,nonacm]{acmart}

\usepackage{amsmath}
\usepackage{amsfonts}
\usepackage{siunitx}

\usepackage[frozencache=true,cachedir=minted]{minted}
\setminted[pycon]{frame=lines,framesep=2mm,linenos,python3=true,fontsize=\small}

\usepackage{acronym}
\acrodef{CDF}{Cumulative Distribution Function}
\acrodef{CMF}{Cumulative Mass Function}
\acrodef{PDF}{Probability Density Function}
\acrodef{PRNG}{Pseudo-Random Number Generator}
\acrodef{AWS}{Amazon Web Services}
\acrodef{GPU}{Graphical Processing Unit}
\acrodef{AES}{Advanced Encryption Standard}
\acrodef{RNG}{Random Number Generator}
\acrodef{PRNG}{Pseudorandom Number Generator}
\acrodef{CSPRNG}{Crytographically Secure Pseudorandom Number Generator}
\acrodef{CPRNG}{Crytographic Pseudorandom Number Generator}
\acrodef{CBPRNG}{Counter-Based Pseudorandom Number Generator}
\acrodef{MT}{Mersenne Twister}
\acrodef{DP}{Differential Privacy}
\acrodef{PCG}{Permuted Congruential Generator}

\def\ie{\textit{i.e.}}
\def\cf{\textit{cf.}}
\def\eg{\textit{e.g.}}
\def\etc{\textit{etc.}}

\def\CC{{C\nolinebreak[4]\hspace{-.05em}\raisebox{.13ex}{++}}}

\newcommand{\secabs}[1]{}
\newcommand{\cfootnote}[1]{}
\newcommand{\ttt}[1]{\colorbox{white!93!black}{\texttt{#1}}}

\begin{document}

\title{Random Number Generators and Seeding for Differential Privacy}

\author{Naoise Holohan}
\email{naoise@ibm.com}
\affiliation{%
\institution{IBM Research Europe -- Ireland}
\city{Dublin}
\country{Ireland}}
\authorsaddresses{}

\begin{abstract}
\ac*{DP} relies on random numbers to preserve privacy, typically utilising \acp*{PRNG} as a source of randomness. In order to allow for consistent reproducibility, testing and bug-fixing in \acs*{DP} algorithms and results, it is important to allow for the seeding of the \acsp*{PRNG} used therein.

In this work, we examine the landscape of \acp*{RNG}, and the considerations software engineers should make when choosing and seeding a \acs*{PRNG} for \acs*{DP}. We hope it serves as a suitable guide for \acs*{DP} practitioners, and includes many lessons learned when implementing seeding for diffprivlib.
\end{abstract}

\hypersetup{
  allcolors = green!50!black,
  urlcolor = red!50!black,
}

\maketitle

\section{Introduction}
The generation of random numbers on computers typically boils down to two distinct steps performed in series~\cite{Lec12}, (i) \emph{bit generation}, generating a stream of random bits that can be coerced into uniform variates in $[0,1)$ or uniformly distributed integers~\cite{Knu97,Lec94,Lec98,Nie92,Tez95}, and (ii) \emph{bit transformation}, the transformation of these uniform variates to generate random variates from arbitrary distributions~\cite{Dev86,Gen03,HLD04}.

In this paper, we are concerned with the first of these steps, bit generation, and the considerations engineers and scientists must make when designing and implementing \ac{DP} systems. Issues around the latter, bit transformation, are beyond the scope of this paper, but are independently of great interest to the \ac{DP} community, because of the privacy vulnerabilities that have recently been highlighted~\cite{HB21,HDH22,JMR22}. \ac{DP}, a popular paradigm for privacy-enhancing computations, uses random noise to preserve privacy, and therefore requires care and attention when generating random numbers.

Among the requirements for a good \ac{RNG} is that it is portable and reproducible to allow for reliable and consistent testing, debugging and reproduction of any results~\cite{Hel98,MWK07}.\footnote{For the purpose of this paper, we refer to a \acf{RNG} as the component performing the bit generation step only, and not bit transformation.} These qualities are also of importance to the \ac{DP} community. Although sensitive \ac{DP} releases for publication ``in the wild'' necessitate the most secure type of randomness -- that produced by a \ac{CSPRNG}~\cite{KMR20} -- non-sensitive \ac{DP} simulations (\ie, for publication, testing and debugging) typically require a user to be able to reproduce results, and therefore require a seeded \ac{RNG}.

In this paper, we examine the elements of choosing, implementing and securing an \ac{RNG} for differential privacy. This includes a introduction to \acp{RNG} (Section~\ref{sc:rng}) and typical pitfalls encountered when seeding (Section~\ref{sc:seed}). We largely focus on implementations using the Python programming language, and the Numpy and Scikit-Learn packages (Section~\ref{sc:python}), with more detailed notes on choices we made for  diffprivlib~\cite{HBML19} (Section~\ref{sc:diffprivlib}). Nevertheless, we hope the principles are sufficiently covered to allow the reader adapt them to their chosen language(s) and package(s). Additional details are provided by way of footnotes, for those interested in delving deeper.

\section{An overview of the Random Number Generator}\label{sc:rng}

Humans have been seeking sources of randomness for millennia. Long before the advent of modern statistics and probability, randomness has been used in determining fate, playing games and selecting public officials. The humble dice can trace its origins back c.~\num{5000} years~\cite{ABE08}, while the notion of tossing coins has been considered since as far back as c.~\num{3000} years ago in the \emph{I~Ching}.

In more recent times, human computers relied on tables of random numbers for their tasks, a well-known example being the book of one million random digits from the appropriately-named RAND Corporation~\cite{Rand55}.\footnote{A valued contribution given how non-random humans tend to be~\cite{KP76}.} Nowadays, randomness has a wide variety of uses across science, art, gaming and other fields, necessitating the requirement to generate vast quantities of random numbers. The US Census Bureau estimated needing 90TB of random bytes to implement differential privacy on its 2020 Census release~\cite{GL20}.

\subsection{Hardware}
Hardware \acp{RNG} are commonly used for critical tasks (such as cryptography), where the supposed randomness of nature is exploited to generate random numbers. Examples include quantum states~\cite{MYC16}, cameras and microphones on mobile devices~\cite{KSM07,NCP03} and walls of lava lamps~\cite{NMS98}\footnote{\url{https://www.youtube.com/watch?v=1cUUfMeOijg}}. On Unix-like operating systems, physical sources including ``inter-keyboard timings'' are used to fill pools of entropy for generating randomness in \ttt{/dev/random} (and \linebreak\ttt{/dev/urandom}, which is preferred by many)\footnote{See \url{https://www.2uo.de/myths-about-urandom/} and \linebreak\url{https://sockpuppet.org/blog/2014/02/25/safely-generate-random-numbers/}.}.\footnote{\url{https://git.kernel.org/pub/scm/linux/kernel/git/stable/linux.git/tree/drivers/char/random.c?id=refs/tags/v3.15.6\#n52}} A discussion on whether these, and other sources, are truly random and unpredictable (\cf\ superdeterminism) is certainly beyond the scope of this paper!

\subsection{Algorithmic}
Given the ubiquity of randomness in science, a more flexible solution to randomness beyond hardware generation is desirable, not least for our ability as scientists to predictably test programmes we write (the very subject of this paper). Around the same time as RAND was producing its book of random digits, John von Neumann was working on the first \ac{PRNG}, predicting: ``When random numbers are to be used in fast machines, number will usually be needed faster.'' Von Neumann's \emph{middle square} method~\cite{VN51} is only of historical interest now, but is still used as inspiration for new \acp{PRNG}~\cite{Wid22}.

Since von Neumann's efforts, the search has continued for fast, memory-efficient and more random \acp{PRNG}. The \ac{MT}~\cite{MN98} is commonly deployed in many programming languages and libraries (including Python and R), while Numpy now uses the \ac{PCG}~\cite{ON14} as its default \ac{PRNG}. There are a host of other \acp{PRNG} available~\cite{Lom08,BD22}.

\subsection{Statistical Testing}
A large catalogue of statistical tests exist to vet the randomness of \acp{RNG} (both hardware and algorithmic). Knuth was the first to devise a series of tests for \acp{RNG}~\cite{Knu97}, and at present there are three libraries\slash batteries of tests that are in common use: NIST STS~\cite{RSN01}, Dieharder~\cite{BEB14} (an update on the original Diehard~\cite{Mar95}, now containing \num{110} tests) and TestU01~\cite{LES07} (which includes the state-of-the-art \emph{Big Crush} collection of \num{106} tests). \acp{RNG} are not expected to be able to pass all tests~\cite{LON19}, and even the tests themselves are subject to rigorous testing~\cite{SOM22}.

Of course, the tests tell nothing of the true randomness of the stream provided for testing. Common mathematical constants such as $\pi$, $e$ and $\sqrt{2}$ are able to pass these tests despite obviously not being random~\cite{Mar05}, as are most rationals $\frac{k}{p}$ for large primes $p$.

\subsection{\aclp{CSPRNG}}
A \ac{PRNG}'s ability to pass statistical tests for randomness (such as BigCrush) may not make it suitable for certain tasks, however. While non-critical tasks, such as Monte Carlo simulation, may only require fast random number generation that passes Big Crush to be suitable, applications in cryptography and privacy may require further guarantees. \acp{RNG} need to pass two additional tests to be a \acf{CSPRNG} (or \ac{CPRNG}): (i) that the next bit cannot be predicted from knowing the previous bits (the \emph{next-bit test}, or \emph{polynomial-time perfection}~\cite{Lec12}), and (ii) that the state of the \ac{RNG} cannot be reverse-engineered from its outputs.

Commonly-used \acp{PRNG} are susceptible to state recovery by observing a sufficiently large sequence of output bits. Examples include the \ac{MT} (whose state can be determined by observing \num{624} consecutive outputs~\cite{AK12}) and \ac{PCG} (for which attacks taking \num{20000} CPU hours have compromised its state~\cite{BMS20}). Hardware \acp{RNG} are typically robust from these state recovery attacks, since they lack any internal state and instead rely on the state of the universe for its randomness.\footnote{For the benefit of mankind, we assume the state of the universe to be unknowable.}

\acp{CSPRNG} need not be used for every application however, since the added security they provide usually comes at the expense of computation time. Examples from the Rust programming language show \acp{CSPRNG} to be up to \num{200} times slower per byte of data than non-cryptographic variants,\footnote{\url{https://docs.rs/nanorand/latest/nanorand/}} although other examples show a difference of just 50\%.\footnote{\href{https://rust-random.github.io/book/guide-rngs.html\#basic-pseudo-random-number-generators-prngs}{https://rust-random.github.io/book/guide-rngs.html}}

A common example of a \ac{CSPRNG} is the AES block cipher in counter mode, which has been in use for two decades~\cite{HK03}. More recently, this idea of using counter-mode encryption protocols as \acp{RNG} has heralded a new class of generator, the \ac{CBPRNG}~\cite{SMD11}. The simple state space of these \acp{RNG} (just a counter and a key) is combined with simplifications in the cryptograpic guarantees (\eg, 5-7 tranformation rounds in ADS instead of the 10-14 rounds in AES) to construct a fast (albeit non-crypto-secure) \ac{PRNG} suitable for simple parallelisation.

Other popular \acp{CSPRNG} include ChaCha~\cite{Ber08} and HC-128.

\subsection{Relevance to \acl{DP}}
When publishing \ac{DP} outputs, it is preferable to use a \ac{CSPRNG} to ensure maximum security against reverse engineering of the noise. For the purpose of testing, scientific publication and other controlled analysis where privacy of the underlying data is not of concern, it is sufficient to use a well-tested \ac{PRNG} given the advantages in availability, execution time and reproducibility. The statistical quality of their randomness is taken care of by the statistical tests, as distinct from the cryptographic tests \acp{CSPRNG} must pass.

\section{Random Number Generation and seeding in Python and Numpy}\label{sc:python}
\secabs{How are random numbers generated?}
We now give an overview of the specifics of \acp{RNG}, and their seeding, in Python and Numpy, given that they are the language and primary library used for building diffprivlib. While the specific implementation of \acp{RNG} will differ with other languages and libraries, we hope this outline will give sufficient principles-based insights to be adaptable to the reader's preferred tools.

\subsection{\texttt{random} module}
The default package for random number generation in Python is the Random module.\footnote{\url{https://docs.python.org/3/library/random.html}} Random uses the \ac{MT} \ac{PRNG} as its core generator (with a period of $2^{19937}-1$), and is invoked by instantiating a generator (\ttt{random.Random()}) with which sampling procedures can be executed (\eg, \ttt{random()}, \ttt{randint()} and \ttt{normalvariate()}).  Alternatively, the sampling procedures can be called directly from module methods (\eg, \ttt{random.randint()}).\footnote{In this case, the generator instance is stored in the \ttt{random.\_inst} variable, so \ttt{random.random()} is equivalent to \ttt{random.\_inst.random()}, \etc}

The generator can be seeded with the \ttt{seed} method (\ie, \ttt{random.Random().seed()}), which sets the state of the \ac{MT}. If no seed is provided (or a \ttt{None} seed), the generator is seeded with operating system randomness (\ie, \ttt{/dev/random}), or, otherwise, the current time and process ID.\footnote{\url{https://github.com/python/cpython/blob/3.11/Modules/\_randommodule.c\#L284-L290}} This ensures pseudorandomness without user-driven seeding, in contrast to other languages such as C which are not seeded by default (meaning repeated execution of scripts with \ttt{rand} will give the same outputs).\cfootnote{\url{https://linuxhint.com/rand-function-in-c-language/}}

\subsection{\texttt{secrets} module}\label{sc:secrets}
Arising from the weak security guarantees of the \ac{MT} \ac{RNG} and the apparent misuse of the module in security-critical tasks, the Secrets module was added to Python in 2015 as a user-friendly \ac{CSPRNG} using \ttt{/dev/random} (in Unix) or \ttt{CryptGenRandom()} (in Windows).\footnote{\url{https://peps.python.org/pep-0506/}} As this module uses OS-generated randomness, it is neither necessary nor possible to seed the generator. However, on most machines, it is the most secure way to generate random bits and should be preferred for security-critical tasks. However, for tasks require reproducible results, Secrets is not suitable.

\subsection{Numpy}\label{sc:numpy}
Numpy is a popular library for scientific computing in Python, supporting calculations on large multidimensional arrays and an extensive library of mathematical functions~\cite{WCV11}. Prior to Numpy~v1.17, \ttt{RandomState} was the class for generating randomness, containing both bit generation (using the \ac{MT}, like Python) and bit transformation functionality. Each of the bit transformers (\eg, \ttt{random\_integers()} and \ttt{standard\_normal()}) can also be called directly from the \ttt{numpy.random} module.\footnote{These are called from the pre-initialised instance of \ttt{RandomState} at \ttt{numpy.random.\_rand}.}

Since Numpy~v1.17,\footnote{See \href{https://numpy.org/doc/1.25/release/1.17.0-notes.html}{release notes for v1.17}, corresponding \href{https://numpy.org/neps/nep-0019-rng-policy.html}{NEP~19} and pull request \href{https://github.com/numpy/numpy/pull/13163}{GH13163}.} the bit generation and bit transformation functionality has been divided between the \ttt{BitGenerator} and \ttt{Generator} classes respectively. This allows for the implementation of different \acp{RNG} that subclass the \ttt{BitGenerator} class, while leveraging the common distribution sampling functionality of \ttt{Generator}. As of Numpy~v1.25, five bit generators are natively supported, including \ac{MT} and \ac{PCG}. A third-party package, \emph{randomgen}, having started as a project to modernise Numpy's Random module, has a number of additional Numpy-compatible \acp{PRNG} and \acp{CSPRNG}.\footnote{\url{https://github.com/bashtage/randomgen}}

When a bit generator is seeded by the user, that seed is used to initialise a new \ttt{SeedSequence} object.\footnote{\url{https://github.com/numpy/numpy/blob/v1.25.0/numpy/random/bit\_generator.pyx\#L533}} A seed sequence mixes the entropy of the provided seed(s) (through iterative hashing and mixing),\footnote{Numpy's implementation (\url{https://github.com/numpy/numpy/pull/13780}) is derived from Melissa E. O'Neill's design (\url{https://www.pcg-random.org/posts/developing-a-seed_seq-alternative.html}).} allowing the \ttt{BitGenerator} to request the entropy required to seed itself (\eg, \num{624} \num{32}-bit integers for \ac{MT}, or two \num{64}-bit integers for \ac{PCG}).
There is no limit to the size of integer a seed sequence can be seeded with.\footnote{Within Numpy, \ttt{SeedSequence} handles a large integer seeds by splitting them into an array of \num{32}-bit Numpy integers (lowest bits first -- with \ttt{\_int\_to\_uint32\_array()}), which is then used to generate an \ac{RNG}'s state. System memory is therefore the only limiting factor in the initial seed size.} If a seed is not explicitly provided, \num{128} random bits are drawn from Secrets to be used as the seed.\footnote{See \url{https://github.com/numpy/numpy/blob/v1.25.0/numpy/random/bit_generator.pyx\#L302}.}

\subsection{Other Python libraries}
Scipy, another popular Python library for scientific computing, uses Numpy for underlying randomness (with support for both \ttt{RandomState} and \ttt{Generator}) and adds extra functionality with its \emph{stats} module~\cite{scipy}.\footnote{\url{https://docs.scipy.org/doc/scipy-1.11.1/reference/stats.html}} TensorFlow, a library primarly used for the training and inference of deep neural networks, has a standalone \ac{RNG} module, with support for two \acp{CBPRNG}~\cite{Tensorflow,SMD11}.\footnote{\url{https://www.tensorflow.org/versions/r2.12/api\_docs/python/tf/random}}

\section{Seeding pitfalls}\label{sc:seed}
\secabs{Cutting edge approach to seeding}
Given humans' inability to select digits uniformly at random, it should come as no surprise that we as a species has a special affinity with certain integers as seeds for \acp{RNG}. A number of studies\footnote{\url{https://www.kaggle.com/code/residentmario/kernel16e284dcb7}, \linebreak\url{https://blog.semicolonsoftware.de/the-most-popular-random-seeds/} and \linebreak\url{https://www.residentmar.io/2016/07/08/randomly-popular.html}} have shown the distribution of seeds chosen ``randomly'' weighs heavily on \num{0}, \num{1}, \num{42}\footnote{The Answer to the Ultimate Question of Life, the Universe, and Everything: \url{https://en.wikipedia.org/wiki/Phrases_from_The_Hitchhiker\%27s_Guide_to_the_Galaxy}.} and \num{1234}\footnote{The kind of combination an idiot would have on their luggage: \url{https://www.youtube.com/watch?v=a6iW-8xPw3k}.}. \CC\ users appear especially attached to the seed \num{380843}, although this author, like others, was unable to get to the bottom of its origins.\cfootnote{Neither Google, Wikipedia nor the OEIS offered any insight of note}

\subsection{Seeding for security and bias mitigation} \label{sc:seed:bias}
\secabs{How do we avoid the pitfalls mentioned above?}
Seeding our \acp{RNG} even with 32-bit integers (up to \num{4294967296}) presents an immediate security risk, even if a \ac{CSPRNG} is being used, as it opens the possibility of the seed being determined by brute force.

Aside from security risks, seeding with fewer bits that the internal state risk bias in the random numbers generated. This may affect the randomness of the \ac{RNG} and have an unanticipated effect in tests and experiments. For example, using an MT19937 \ac{RNG} to sample \num{32}-bit integers, it will never produce \num{7} nor \num{3} as the first output if it were seeded with a single \num{32}-bit integer!\footnote{\url{https://www.pcg-random.org/posts/cpp-seeding-surprises.html}} Since the internal state of an MT19937 is \num{624} \num{32}-bit integers, it should be no surprise that seeding with only \num{32} bits will leave holes in its output space. Other defects of seeded \acp{RNG} have been found through simulations of baseball games~\cite{MWK07}.

\paragraph{Numpy} The use of \ttt{SeedSequence} is advised in Numpy (see Section~\ref{sc:numpy}). It is recommended to initialise a \ttt{SeedSequence} object without a seed (for which the Secrets package is used to extract randomness) and a seed can then be extracted via the \ttt{entropy} attribute to generate a static seed for use in downstream tasks (see Listing~\ref{lst:seedseq}).\footnote{\url{https://numpy.org/doc/1.25/reference/random/bit_generators/generated/numpy.random.SeedSequence.html}} This should reduce the chance of bias in results, and eliminate it in the case of \acp{RNG} that have a 128-bit internal state (\ie, \ac{PCG}).

\begin{listing}
\begin{minted}{pycon}
>>> from numpy.random import SeedSequence, PCG64, Generator
>>> SeedSequence().entropy
287955962967732827663192315245491885249
>>> rng = Generator(PCG64(287955962967732827663192315245491885249))
>>> rng.random()
0.07296271584154868
\end{minted}
\caption{Using \ttt{SeedSequence} to seed an \ac{RNG} in Numpy.}
\label{lst:seedseq}
\end{listing}

\subsection{Seeding for parallel processes}
\secabs{What's the best method for seeding in parallel? Sharing states not possible?}
The generation of random numbers in parallel applications has been an area of active research for decades~\cite{DMP88,Hel98,SMD11,LMO17},\cfootnote{Others include~\cite{APG92,Cod97,Hel98,LBC93,MS00,DMP88,Tan02}} and requires considerable caution to be observed. There are a number of techniques for producing parallel (sub-)streams, the most common of which are (i) single stream, (ii) multi-stream, and (iii) substream~\cite{LMO17,SMD11}.

Using a single stream is not advised, due to speed and reproducibility limitations. Multistreaming can be achieved by parametrising a local stream for each computational unit. This can be done deterministically (\ie, using a thread ID as a key for a \ac{CBPRNG}), or by using a randomly generated seed. Random seeding is particularly useful when generators must be created on-the-fly (although care must be taken to avoid the ``birthday problem''~\cite{McK66} of re-using the same stream). In most situations, the probability of overlapping streams is negligible~\cite{LMO17}.

Substreaming can be achieved in two ways. With \emph{leap-frogging} (or \emph{striding}), task $t\in [T]$ gets all integers $n_i$ satisfying $i \mod T = t$ (\ie, every $T$th sample).  Either each task is required to consume one value at each step, or each task can advance its own generator by $T$ steps each time, a typically time consuming process. Alternatively, with \emph{blocking}, task $t$ gets $N$ integers $n_i$ where $i \in [t N, (t + 1)N - 1]$. This requires the \ac{RNG} to have a jump-ahead function, which is not universal. Care must also be taken that the same blocks are not re-used, particularly when iteratively or adaptively blocking (\ie, a grandchild sharing the same block as their aunt).

\paragraph{Numpy} The \ttt{SeedSequence} object is again recommended to generate seeds for parallel generation in Numpy,\footnote{\url{https://numpy.org/doc/1.25/reference/random/parallel.html}} taking advantage of the uniqueness of outputs to avoid the birthday problem. It is also an explicit, user-friendly way to seed the parallel generators, and should be easy to follow from a cursory reading of the code. With \ttt{SeedSequence}, new seeds can be generated from the child \ac{RNG}, allowing for grandchild and great-grandchild processes to be spawned without having to pre-allocate space and without running the risk of overlapping streams.\footnote{Each seed sequence is given a unique key that tracks its position in the tree of sequences spawned from the root (\ie, the first grandchild from a first child sequence is given the key \ttt{(0,0)}) while using the same entropy. The \ttt{spawn\_key} is mixed with the entropy pool to create a unique pool of randomness for each spawned \ac{RNG}: \url{https://github.com/numpy/numpy/blob/v1.25.0/numpy/random/bit\_generator.pyx\#L308-L313}.}

Since Numpy~v1.25, a \ttt{Generator} class instance can be used to spawn child generators for parallel computation using the \ttt{spawn()} method,\footnote{\url{https://github.com/numpy/numpy/pull/23195}} while older versions of Numpy can use the \ttt{BitGenerator}'s \ttt{\_seed\_seq} attribute to access the \ac{RNG}'s underlying seed sequence (which can be used to spawn new sequences with which to initialise new generators).\footnote{\url{https://github.com/numpy/numpy/issues/15322\#issuecomment-626400433}}

Numpy also supports blocking of some \acp{RNG} using the \ttt{jumped()} method (spawning a new \ac{RNG} whose state has been jumped as if $2^{128}$ -- or $(\phi-1)2^{128}$ for \ac{PCG} -- random numbers have been drawn).

\section{Seeding in diffprivlib}\label{sc:diffprivlib}
\secabs{What do we need to consider when adding seeding in DP?}
We have two priorities in implementing seeded random number generation in diffprivlib. Firstly, to maintain a consistent, user-friendly interface for seeding across all diffprivlib functions, and equally importantly, the ability to use a cryptographically secure source of randomness for DP noise generation (\eg, using Secrets) in sensitive tasks. Thankfully, precedent exists in both Numpy (Section~\ref{sc:numpy}) and Scikit-Learn for seeding, and leveraging this has simplified the engineering challenge for diffprivlib.

Support for \ac{RNG} seeding was added in diffprivlib v0.6.\footnote{\url{https://github.com/IBM/differential-privacy-library/releases/tag/0.6.0}}

\subsection{Lessons from Scikit-Learn and Numpy}
\secabs{Use a single \ac{RNG} instance, seeding with an interger vs with an instance; rewrite with a language-independent approach}
The core philosophy behind \ac{RNG} seeding in Numpy and Scikit-Learn is the ability to pass an single \ac{RNG} instance to a function\slash method, and have it persist and propagate through all calls and sub-calls within its logic. This \ac{RNG} instance is what governs all elements of randomness within that function\slash method, whether specifically relied upon for \ac{DP} or not. This ensures the user has full control over the randomness of a particular function, and the randomness of different function calls within the same script, if required.\footnote{As described in more detail by Numpy developer Robert Kern (see thread of comments): \url{https://stackoverflow.com/a/5837352}.} This provides more flexibility and control, and (importantly for this paper) gives an explicit method for seeding that should be clear from a cursory reading of the code. It also allows for multiple streams to be maintained without cross-contamination. This is also consistent with established software engineering practices of prioritising local variables over globals~\cite{MC09}, as they can reduce flexibility and modularity of code.

In terms of implementation, each diffprivlib function takes a \ttt{random\_state} parameter which can be (i) \ttt{None}, (ii) a non-negative integer, or (iii) a \ttt{RandomState} instance. The \ttt{random\_state} parameter is ingested by the \ttt{check\_random\_state()} function. When a \ttt{RandomState} instance is provided, that instance is used as the generator, updating its state when deployed.

If the provided random state is an integer, \ttt{check\_random\_state()} creates a new \ttt{RandomState} instance, seeded with that integer. If no seed is provided (or \ttt{seed=None}), the \linebreak\ttt{numpy.random.mtrand.\_rand} instance is returned. The returned instance is then used for all calls within the function, and any calls derived from it.

There is an important distinction between providing integer-valued seeds and \ttt{RandomState} instances that are worth highlighting.\footnote{Further details are available here: \url{https://scikit-learn.org/1.2/common\_pitfalls.html\#controlling-randomness}.} Integer-valued seeds are a great way to produce the same outputs for repeat calls of the same function. Consider the code using a (poorly-chosen, per Section~\ref{sc:seed}) integer-valued seed in Listing~\ref{lst:int} and the corresponding output when parametrising with a persistent \ttt{RandomState} instance in Listing~\ref{lst:rng}. In the latter, the function calls are no longer returning the same output, but running the same \emph{script} again will consistently produce the same output. In most cases, this behaviour is what is desired.

\begin{listing}
\begin{minted}{pycon}
>>> from diffprivlib.mechanisms import Laplace
>>> Laplace(epsilon=1, sensitivity=1, random_state=1).randomise(1)
0.6556851745442447
>>> Laplace(epsilon=1, sensitivity=1, random_state=1).randomise(1)
0.6556851745442447
\end{minted}
\caption{Seeding a diffprivlib mechanism with an integer giving the same output.}
\label{lst:int}
\end{listing}

\begin{listing}
\begin{minted}{pycon}
>>> from diffprivlib.utils import check_random_state
>>> rng = check_random_state(1)
>>> Laplace(epsilon=1, sensitivity=1, random_state=rng).randomise(1)
0.6556851745442447
>>> Laplace(epsilon=1, sensitivity=1, random_state=rng).randomise(1)
1.2482045505881658
\end{minted}
\caption{Using an \ac{RNG} instance with a diffprivlib mechanism gives unique (but reproducible) outputs.}
\label{lst:rng}
\end{listing}

\subsection{Limitations with Scikit-Learn}
\secabs{Uses old Numpy RNG, only accepts 32-bit seeds}
There are a number of issues presented by Scikit-Learn's approach to random number generation that limits our adoption of best practices in diffprivlib.

\paragraph{Legacy random state} Firstly, only the legacy \ttt{RandomState} interface is natively supported by Scikit-Learn.\footnote{The Scikit-Learn community is considering supporting Numpy's \ttt{BitGenerator} for randomness (\href{https://github.com/scikit-learn/scikit-learn/issues/16988}{GH16988} and \href{https://github.com/scikit-learn/scikit-learn/issues/20669}{GH20669}) as is Scipy (\href{https://github.com/scipy/scipy/issues/14322}{GH14322}).} This limits the generator to the slower MT19937, and excludes the better \acp{RNG} introduced in Numpy~v1.17 (or any other custom \acp{RNG}). It is possible to work around this \ac{RNG} limitation by explicitly specifying a \ttt{BitGenerator} as a seed for a \ttt{RandomState} instance (\ie, \ttt{np.random.RandomState(np.random.PCG64(seed))}), assuming the required version of Numpy is installed.\footnote{As of Scikit-Learn~v1.1, Numpy~v1.17 is a minimum requirement.} However, at present, there is no method to deploy a \ttt{Generator} instance, and make use of its superior bit transformations.\footnote{Nevertheless, issues for \ac{DP} may persist, including floating-point vulnerabilities~\cite{HB21}.}

\paragraph{Low-bit seeds} Another downside to Scikit-Learn's present approach to seeding is its limitation to \num{32}-bit seeds.\footnote{Not only are seeds larger than \num{32} bits unsupported \linebreak(\url{https://github.com/scikit-learn/scikit-learn/blob/1.3.0/sklearn/utils/\_param\_validation.py\#L548}), but supplying one is met with an uninformative \ttt{TypeError} (a fix for which is in progress at time of writing, \href{https://github.com/scikit-learn/scikit-learn/pull/26648}{GH26648}).} This limitation appears to be arbitrary, since the seed is directly passed to a \ttt{RandomState} instance which routinely accepts larger integer seeds (that are passed straight through to a \ttt{SeedSequence} instance to generate the required randomness bit-count). Additionally, it is at odds with Numpy's preferred practice of using the entropy of a \ttt{SeedSequence}, itself a \num{128}-bit integer. A workaround to this is again to pass the seed (or seed sequence) directly to the \ttt{BitGenerator}, which can then be passed to the \ttt{RandomState}. It is planned to add this workaround to diffprivlib in due course.

\subsection{Deploying Secrets}
\secabs{Want secrets used in mechanisms whenever a seed isn't supplied}
As outlined in Section~\ref{sc:secrets}, the Secrets module is the best source of cryptographically-secure randomness in Python, thanks to its hardware-generated pools of randomness entropy. From a privacy perspective, the only application of randomness that needs to be secured is that relating to DP, although care must be given to other aspects of randomness which may be relied upon for DP (\eg, privacy amplification through shuffling~\cite{EFM19}). In diffprivlib, most of the important DP noise additions happen within the \emph{mechanisms} module.

For reproducibility purposes, we must use the relevant \ttt{RandomState} instance whenever a seed has been supplied, but use Secrets with diffprivlib-specific components if a seed hasn't been supplied. This can be done with relative ease because the default \ttt{RandomState} returned when \ttt{seed=None} is the one stored at \ttt{numpy.random.mtrand.\_rand};\footnote{The stability of this feature for Scikit-Learn now seems to be considered a mistake: \url{https://numpy.org/neps/nep-0019-rng-policy.html\#numpy-random}.} if this is the same as the supplied \ac{RNG} to the mechanism, we know to override the \ac{RNG} with Secrets.

One final consideration when allowing for secrets to be used on-the-fly are the differences in sampling function names. The interface of \ttt{SystemRandom} (as used by Secrets) differs from that used by \ttt{RandomState}, requiring the use of \ttt{try...except} statements wherever a difference occurs (thankfully, both interfaces employ \ttt{random()} for uniform variate sampling).

\section{Conclusion and TL;DR}
\acf{DP} practitioners are encouraged to support the seeding of \acfp{RNG} to allow for reproduction of any published work (using open, non-sensitive data), and to help with testing and debugging. Choosing any rigorously-tested \acf{PRNG} is likely to be sufficient, but special care must be taken when seeding (especially for parallel computations).

Be sure to seed the \ac{RNG} with many random bits to avoid biased outcomes. When seeding \acp{RNG} for parallel computations, use a seed sequence if possible (they have many benefits, including that they are bijective, and can spawn non-overlapping streams with ease). Otherwise, block the \ac{RNG} (using a jump-ahead function like \ttt{jumped()}) while keeping track of previously-assigned blocks to avoid overlaps.

Lastly, to enable safe \ac{DP} releases ``in the wild'', allow for the use of a \acf{CSPRNG} for the secure generation of noise whenever a seed is not provided.

\begin{acks}
This work was supported by European Union's Horizon 2020 research and innovation programme under grant number 951911 -- AI4Media.
\end{acks}

\bibliographystyle{acmalpha}
\bibliography{refs}

\newcommand{\etalchar}[1]{$^{#1}$}
\begin{thebibliography}{HBMAL19}

\bibitem[AAB{\etalchar{+}}15]{Tensorflow}
{\sc Abadi, M., Agarwal, A., Barham, P., Brevdo, E., Chen, Z., Citro, C.,
  Corrado, G.~S., Davis, A., Dean, J., Devin, M., Ghemawat, S., Goodfellow, I.,
  Harp, A., Irving, G., Isard, M., Jozefowicz, R., Jia, Y., Kaiser, L., Kudlur,
  M., Levenberg, J., Mané, D., Schuster, M., Monga, R., Moore, S., Murray, D.,
  Olah, C., Shlens, J., Steiner, B., Sutskever, I., Talwar, K., Tucker, P.,
  Vanhoucke, V., Vasudevan, V., Viégas, F., Vinyals, O., Warden, P.,
  Wattenberg, M., Wicke, M., Yu, Y., and Zheng, X.}
\newblock Tensor{F}low, large-scale machine learning on heterogeneous systems,
  Nov. 2015.

\bibitem[ABE08]{ABE08}
{\sc Aruz, J., Benzel, K., and Evans, J.~M.}
\newblock {\em Beyond {B}abylon: art, trade, and diplomacy in the second
  millennium {BC}}.
\newblock Metropolitan Museum of Art, 2008.

\bibitem[AK12]{AK12}
{\sc Argyros, G., and Kiayias, A.}
\newblock I forgot your password: Randomness attacks against {PHP}
  applications.
\newblock In {\em USENIX Security Symposium\/} (2012), pp.~81--96.

\bibitem[BD22]{BD22}
{\sc Bhattacharjee, K., and Das, S.}
\newblock A search for good pseudo-random number generators: Survey and
  empirical studies.
\newblock {\em Computer Science Review 45\/} (2022), 100471.

\bibitem[BEB14]{BEB14}
{\sc Brown, R.~G., Eddelbuettel, D., and Bauer, D.}
\newblock Dieharder: A random number test suite (version 3.31.1).
\newblock {\em \url{http://www.phy.duke.edu/\~rgb/General/dieharder.php}\/}
  (2014), 1.

\bibitem[Ber08]{Ber08}
{\sc Bernstein, D.~J.}
\newblock Cha{C}ha, a variant of {S}alsa20.
\newblock In {\em Workshop record of SASC\/} (2008), vol.~8, Citeseer,
  pp.~3--5.

\bibitem[BMS20]{BMS20}
{\sc Bouillaguet, C., Martinez, F., and Sauvage, J.}
\newblock Practical seed-recovery for the {PCG} pseudo-random number generator.
\newblock {\em IACR Transactions on Symmetric Cryptology\/} (2020).

\bibitem[Dev86]{Dev86}
{\sc Devroye, L.}
\newblock {\em Non-uniform random variate generation}.
\newblock Springer-Verlag, New York, 1986.

\bibitem[DMP88]{DMP88}
{\sc De~Matteis, A., and Pagnutti, S.}
\newblock Parallelization of random number generators and long-range
  correlations.
\newblock {\em Numerische Mathematik 53\/} (1988), 595--608.

\bibitem[EFM{\etalchar{+}}19]{EFM19}
{\sc Erlingsson, {\'{U}}., Feldman, V., Mironov, I., Raghunathan, A., Talwar,
  K., and Thakurta, A.}
\newblock {\em Amplification by Shuffling: From Local to Central Differential
  Privacy via Anonymity}.
\newblock SIAM, 2019, pp.~2468--2479.

\bibitem[Gen03]{Gen03}
{\sc Gentle, J.~E.}
\newblock {\em Random number generation and {M}onte {C}arlo methods}, vol.~381.
\newblock Springer, 2003.

\bibitem[GL20]{GL20}
{\sc Garfinkel, S.~L., and Leclerc, P.}
\newblock Randomness concerns when deploying differential privacy.
\newblock In {\em Proceedings of the 19th Workshop on Privacy in the Electronic
  Society\/} (New York, NY, USA, 2020), WPES'20, Association for Computing
  Machinery, p.~73–86.

\bibitem[HB21]{HB21}
{\sc Holohan, N., and Braghin, S.}
\newblock Secure random sampling in differential privacy.
\newblock In {\em Computer Security - {ESORICS} 2021 - 26th European Symposium
  on Research in Computer Security, Darmstadt, Germany, October 4-8, 2021,
  Proceedings, Part {II}\/} (2021), E.~Bertino, H.~Shulman, and M.~Waidner,
  Eds., vol.~12973 of {\em Lecture Notes in Computer Science}, Springer,
  pp.~523--542.

\bibitem[HBMAL19]{HBML19}
{\sc Holohan, N., Braghin, S., Mac~Aonghusa, P., and Levacher, K.}
\newblock Diffprivlib: the {IBM} differential privacy library.
\newblock {\em ArXiv e-prints 1907.02444 [cs.CR]\/} (July 2019).

\bibitem[HDH{\etalchar{+}}22]{HDH22}
{\sc Haney, S., Desfontaines, D., Hartman, L., Shrestha, R., and Hay, M.}
\newblock Precision-based attacks and interval refining: how to break, then
  fix, differential privacy on finite computers.
\newblock {\em ArXiv e-prints 2207.13793 [cs.CR]\/} (2022).

\bibitem[Hel98]{Hel98}
{\sc Hellekalek, P.}
\newblock Don't trust parallel {M}onte {C}arlo!
\newblock In {\em Proceedings of the Twelfth Workshop on Parallel and
  Distributed Simulation\/} (USA, 1998), PADS '98, IEEE Computer Society,
  p.~82–89.

\bibitem[HLD04]{HLD04}
{\sc H\"ormann, W., Leydold, J., and Derflinger, G.}
\newblock {\em Automatic nonuniform random variate generation}.
\newblock Statistics and Computing. Springer, 2004.

\bibitem[HW03]{HK03}
{\sc Hellekalek, P., and Wegenkittl, S.}
\newblock Empirical evidence concerning {AES}.
\newblock {\em ACM Trans. Model. Comput. Simul. 13}, 4 (oct 2003), 322–333.

\bibitem[JMRO22]{JMR22}
{\sc Jin, J., McMurtry, E., Rubinstein, B. I.~P., and Ohrimenko, O.}
\newblock Are we there yet? {T}iming and floating-point attacks on differential
  privacy systems.
\newblock In {\em 2022 IEEE Symposium on Security and Privacy (SP)\/} (2022),
  pp.~473--488.

\bibitem[KMR{\etalchar{+}}20]{KMR20}
{\sc Kifer, D., Messing, S., Roth, A., Thakurta, A., and Zhang, D.}
\newblock Guidelines for implementing and auditing differentially private
  systems.
\newblock {\em ArXiv e-prints 2002.04049 [cs.CR]\/} (Feb. 2020).

\bibitem[Knu97]{Knu97}
{\sc Knuth, D.~E.}
\newblock {\em The Art of Computer Programming, Volume 2 (3rd Ed.):
  Seminumerical Algorithms}.
\newblock Addison-Wesley Longman Publishing Co., Inc., USA, 1997.

\bibitem[KP76]{KP76}
{\sc Kubovy, M., and Psotka, J.}
\newblock The predominance of seven and the apparent spontaneity of numerical
  choices.
\newblock {\em Journal of Experimental Psychology: Human Perception and
  Performance 2}, 2 (1976), 291.

\bibitem[K{\v{S}}M07]{KSM07}
{\sc Krhovj{\'a}k, J., {\v{S}}venda, P., and Maty{\'a}{\v{s}}, V.}
\newblock The sources of randomness in mobile devices.
\newblock In {\em Proceedings of the 12th Nordic Workshop on Secure IT Systems.
  NordSec\/} (2007).

\bibitem[L'{E}94]{Lec94}
{\sc L'{E}cuyer, P.}
\newblock Uniform random number generation.
\newblock {\em Annals of Operations Research 53\/} (1994), 77--120.

\bibitem[L'{E}98]{Lec98}
{\sc L'{E}cuyer, P.}
\newblock Random number generation.
\newblock In {\em Handbook of simulation: principles, methodology, advances,
  applications, and practice}, J.~Banks, Ed. Wiley, 1998, pp.~93--137.

\bibitem[L'E12]{Lec12}
{\sc L'Ecuyer, P.}
\newblock {\em Random number generation}.
\newblock Springer, 2012.

\bibitem[LMOS17]{LMO17}
{\sc L’Ecuyer, P., Munger, D., Oreshkin, B., and Simard, R.}
\newblock Random numbers for parallel computers: Requirements and methods, with
  emphasis on {GPU}s.
\newblock {\em Mathematics and Computers in Simulation 135\/} (2017), 3--17.
\newblock Special Issue: 9th IMACS Seminar on Monte Carlo Methods.

\bibitem[LO19]{LON19}
{\sc Lemire, D., and O'Neill, M.~E.}
\newblock Xorshift1024*, xorshift1024+, xorshift128+ and xoroshiro128+ fail
  statistical tests for linearity.
\newblock {\em J. Comput. Appl. Math. 350\/} (2019), 139--142.

\bibitem[Lom08]{Lom08}
{\sc Lomont, C.}
\newblock Random number generation.
\newblock {\em \url{https://lomont.org/papers/2008/Lomont\_PRNG\_2008.pdf}\/}
  (2008).

\bibitem[LS07]{LES07}
{\sc L'Ecuyer, P., and Simard, R.}
\newblock Test{U}01: A {C} library for empirical testing of random number
  generators.
\newblock {\em ACM Trans. Math. Softw. 33}, 4 (aug 2007).

\bibitem[Mar95]{Mar95}
{\sc Marsaglia, G.}
\newblock Diehard battery of tests of randomness.
\newblock {\em
  \linebreak\url{https://web.archive.org/web/20120102192622/www.stat.fsu.edu/pub/diehard}\/}
  (1995).

\bibitem[Mar05]{Mar05}
{\sc Marsaglia, G.}
\newblock On the randomness of $\pi$ and other decimal expansions.
\newblock {\em InterStat 5\/} (2005).

\bibitem[MC09]{MC09}
{\sc Martin, R.~C., and Coplien, J.~O.}
\newblock {\em Clean code: a handbook of agile software craftsmanship}.
\newblock Prentice Hall, Upper Saddle River, NJ [etc.], 2009.

\bibitem[Mc{K}66]{McK66}
{\sc Mc{K}inney, E.~H.}
\newblock Generalized birthday problem.
\newblock {\em The American Mathematical Monthly 73}, 4 (1966), 385--387.

\bibitem[MN98]{MN98}
{\sc Matsumoto, M., and Nishimura, T.}
\newblock Mersenne twister: A 623-dimensionally equidistributed uniform
  pseudo-random number generator.
\newblock {\em {ACM} Trans. Model. Comput. Simul. 8}, 1 (1998), 3--30.

\bibitem[MWKA07]{MWK07}
{\sc Matsumoto, M., Wada, I., Kuramoto, A., and Ashihara, H.}
\newblock Common defects in initialization of pseudorandom number generators.
\newblock {\em ACM Transactions on Modeling and Computer Simulation (TOMACS)
  17}, 4 (2007), 15--es.

\bibitem[MYC{\etalchar{+}}16]{MYC16}
{\sc Ma, X., Yuan, X., Cao, Z., Qi, B., and Zhang, Z.}
\newblock Quantum random number generation.
\newblock {\em npj Quantum Information 2}, 1 (2016), 1--9.

\bibitem[NCP03]{NCP03}
{\sc Noll, L.~C., Cooper, S., and Pleasant, M.}
\newblock What is {L}ava{R}nd.
\newblock {\em \url{http://www.lavarnd.org}\/} (2003).

\bibitem[Nie92]{Nie92}
{\sc Niederreiter, H.}
\newblock {\em Random Number Generation and Quasi-{M}onte {C}arlo Methods}.
\newblock Society for Industrial and Applied Mathematics, 1992.

\bibitem[NMS98]{NMS98}
{\sc Noll, L.~C., Mende, R.~G., and Sisodiya, S.}
\newblock Method for seeding a pseudo-random number generator with a
  cryptographic hash of a digitization of a chaotic system, Mar.~24 1998.
\newblock US Patent 5,732,138.

\bibitem[O'N14]{ON14}
{\sc O'Neill, M.~E.}
\newblock {PCG}: A family of simple fast space-efficient statistically good
  algorithms for random number generation.
\newblock Tech. Rep. HMC-CS-2014-0905, Harvey Mudd College, Claremont, CA,
  Sept. 2014.

\bibitem[{RAN}55]{Rand55}
{\sc {RAND Corporation}}.
\newblock {\em A million random digits with 100,000 normal deviates}.
\newblock Free Press, 1955.

\bibitem[RSN{\etalchar{+}}01]{RSN01}
{\sc Rukhin, A., Soto, J., Nechvatal, J., Smid, M., and Barker, E.}
\newblock A statistical test suite for random and pseudorandom number
  generators for cryptographic applications.
\newblock Tech. rep., Booz-allen and hamilton inc mclean va, 2001.

\bibitem[SMDS11]{SMD11}
{\sc Salmon, J.~K., Moraes, M.~A., Dror, R.~O., and Shaw, D.~E.}
\newblock Parallel random numbers: as easy as 1, 2, 3.
\newblock In {\em Proceedings of 2011 international conference for high
  performance computing, networking, storage and analysis\/} (2011), pp.~1--12.

\bibitem[SOMK22]{SOM22}
{\sc S{\`y}s, M., Obr{\'a}til, L., Maty{\'a}{\v{s}}, V., and Klinec, D.}
\newblock A bad day to die hard: Correcting the dieharder battery.
\newblock {\em Journal of Cryptology 35\/} (2022), 1--20.

\bibitem[Tez95]{Tez95}
{\sc Tezuka, S.}
\newblock {\em Uniform Random Numbers: Theory and Practice}.
\newblock Springer Science+Business Media, New York, 1995.

\bibitem[vdWCV11]{WCV11}
{\sc van~der Walt, S., Colbert, S.~C., and Varoquaux, G.}
\newblock The {N}um{P}y array: A structure for efficient numerical computation.
\newblock {\em Computing in Science and Engineering 13}, 2 (2011), 22--30.

\bibitem[VGO{\etalchar{+}}20]{scipy}
{\sc Virtanen, P., Gommers, R., Oliphant, T.~E., Haberland, M., Reddy, T.,
  Cournapeau, D., Burovski, E., Peterson, P., Weckesser, W., Bright, J., {van
  der Walt}, S.~J., Brett, M., Wilson, J., Millman, K.~J., Mayorov, N., Nelson,
  A. R.~J., Jones, E., Kern, R., Larson, E., Carey, C.~J., Polat, {\.I}., Feng,
  Y., Moore, E.~W., {VanderPlas}, J., Laxalde, D., Perktold, J., Cimrman, R.,
  Henriksen, I., Quintero, E.~A., Harris, C.~R., Archibald, A.~M., Ribeiro,
  A.~H., Pedregosa, F., {van Mulbregt}, P., and {SciPy 1.0 Contributors}}.
\newblock Sci{P}y 1.0: Fundamental algorithms for scientific computing in
  {P}ython.
\newblock {\em Nature Methods 17\/} (2020), 261--272.

\bibitem[VN51]{VN51}
{\sc Von~Neumann, J.}
\newblock Various techniques used in connection with random digits.
\newblock {\em Applied Math Series 12}, 36-38 (1951), 1.

\bibitem[Wid17]{Wid22}
{\sc Widynski, B.}
\newblock Middle-square {W}eyl sequence {RNG}.
\newblock {\em arXiv preprint arXiv:1704.00358\/} (2017).

\end{thebibliography}

\end{document}